\documentclass[12pt,preprint]{aastex}

\slugcomment{\today}

\shorttitle{Hyperon Interaction and Black Hole Formation}
\shortauthors{Nakazato et al.}

\begin{document}


\title{Hyperon Matter and Black Hole Formation in Failed Supernovae}


\author{Ken'ichiro Nakazato\altaffilmark{1}, Shun Furusawa\altaffilmark{2}, Kohsuke Sumiyoshi\altaffilmark{3,4}, Akira Ohnishi\altaffilmark{5}, \\ Shoichi Yamada\altaffilmark{2,6} and Hideyuki Suzuki\altaffilmark{1}}

\email{nakazato@rs.tus.ac.jp}


\altaffiltext{1}{Department of Physics, Faculty of Science \& Technology, Tokyo University of Science, Yamazaki 2641, Noda, Chiba 278-8510, Japan}
\altaffiltext{2}{Department of Physics, Faculty of Science \& Engineering, Waseda University, 3-4-1 Okubo, Shinjuku, Tokyo 169-8555, Japan}
\altaffiltext{3}{Numazu College of Technology, Ooka 3600, Numazu, Shizuoka 410-8501, Japan}
\altaffiltext{4}{Theory Center, High Energy Accelerator Reseach Organization (KEK), Oho 1-1, Tsukuba, Ibaraki 305-0801, Japan}
\altaffiltext{5}{Yukawa Institute for Theoretical Physics, Kyoto University, Kita-shirakawa Oiwake-cho, Sakyo, Kyoto 606-8502, Japan}
\altaffiltext{6}{Advanced Research Institute for Science \& Engineering, Waseda University, 3-4-1 Okubo, Shinjuku, Tokyo 169-8555, Japan}


\begin{abstract}
We investigate the emergence of hyperons in black-hole-forming failed supernovae, which are caused by the dynamical collapse of nonrotating massive stars. We perform neutrino-radiation hydrodynamical simulations in general relativity adopting realistic hyperonic equation-of-state (EOS). Attractive and repulsive cases are examined for the potential of $\Sigma$ hyperons. Since hyperons soften the EOS, they shorten the time interval from the bounce to black hole formation, which corresponds to the duration of neutrino emission. This effect is larger for the attractive case than the repulsive case because $\Sigma$ hyperons appear more easily. In addition, we investigate the impacts of pions to find that they also promotes the recollapse towards the black hole formation. 
\end{abstract}


%% subject header.  Objects should be in the appropriate "individual"
%% headers (e.g. quasars: individual, stars: individual, etc.) with the
%% additional provision that the total number of headers, including each
%% individual object, not exceed six.  The \objectname{} macro, and its
%% alias \object{}, is used to mark each object.  The macro takes the object
%% name as its primary argument.  This name will appear in the paper
%% and serve as the link's anchor in the electronic edition if the name
%% is recognized by the data centers.  The macro also takes an optional
%% argument in parentheses in cases where the data center identification
%% differs from what is to be printed in the paper.



\section{Introduction} \label{intro}

The gravitational collapse of massive stars is the fate of the stellar evolution, which leads to the variety of explosive phenomena, the formation of compact objects and the neutrino burst. The type Ib, Ic and II supernovae are thought to be driven by the core collapse due to photodisintegration reaction of nucleus. The collapse is bounced by the nuclear repulsion and the shock wave is launched invoking explosion. They lead to the neutron stars. In somewhat low mass cases, the core collapse may be triggered by electron captures before Ne ignition, which is called electron-capture supernovae. However, the mass range of these supernova progenitors is still uncertain. If the stellar mass is lower, a white dwarf will be formed finally. \citet{poe08} investigated stellar evolution sequences with initial masses between 6.5 and $13M_\odot$ and the solar metallicity using three different codes and found that the lower mass limit for supernovae is 9-$12M_\odot$. Observationally, \citet{smartt09} found that it converges to $8 \pm 1M_\odot$ from the direct detections of progenitors.

Massive stars beyond a certain mass limit lead to different fates from ordinary supernovae though 
the upper mass limit is more uncertain. \citet{smartt09} suggested that the majority of massive stars above $\sim$20$M_\odot$ may collapse quietly to black holes and that the explosions remain undetected. They investigated 92 samples which complete the core collapse supernovae occurred in a fixed 10.5-year period within a distance of 28 Mpc and concluded that there in no evidence of massive progenitors. Apart from the direct detections, according to the lightcurve models, \citet{nomoto06} proposed that the massive stars beyond 25$M_\odot$ form black holes and their fate splits into two branches, namely, a hypernova branch and a faint-supernova branch. While the constituents of the hypernova branch, which have possible relevance with gamma ray bursts, are suggested to be strongly rotating, those of the faint-supernova branch are nonrotating or weakly rotating.

Having observational suggestions, 
a quiet death through the core collapse occurs for a certain fraction of the massive stars involving collapse. They are thought to become black holes eventually.  Their fate is theoretically proposed to split further into two categories: prompt and delayed black hole formations. While the delayed black hole formation occurs after the faint (perhaps even undetected) supernova explosion, 
the prompt black hole formation occurs when the supernova explosion fails. 
\citet{fryer99} showed that nonrotating stars with the initial mass of $\gtrsim$40$M_\odot$ become failed supernovae according to his numerical simulations. Lately, \citet{ocon10} performed the core collapse simulations for many progenitors with the spherically symmetric models involving simplified neutrino transfer. They found that the threshold between the prompt and delayed black hole formations depends severely on the progenitor models, or evolutionary calculations.

For the investigation of the black hole formation, implementation of properties of dense matter at supranuclear density is mandatory. In particular, meson condensation, hyperon appearance and quark deconfinement are expected, affecting the equation of state (EOS) markedly. For the delayed black hole formations, a proto--neutron star is formed once and recollapse to a black hole $\gtrsim$10~sec after the bounce. \citet{keil95} investigated evolutionary calculations of the proto--neutron star cooling taking into account hyperons. Their result is adopted as the initial condition of the dynamical simulation for the final 100~ms before the delayed black hole formation performed by \citet{baum96}. Incidentally, the evolutions of proto--neutron stars are studied also with kaon condensation \citep[][]{pons01a} and quark deconfinement \citep[][]{pons01b}.  
Those results clarified that the exotic phases appear at late stage 
of the evolution of proto--neutron stars for $\sim$20 sec.  

The prompt black hole formation is a promising branch to explore the exotic phase, 
since the exotic phase appears dramatically right after the core bounce.  
The prompt black hole formation occurs within $\sim$1~sec after the bounce \citep[e.g.,][]{lieb04,sumi06,sumi07} depending on the EOS. 
While no electromagnetic signal other than the disappearance of progenitors is expected for the failed supernova involving the prompt black hole formation \citep[][]{kocha08}, it is as bright in neutrino emissions as ordinary core-collapse supernovae. Recently, numerical simulations of failed supernovae 
utilizing the EOS with hyperons \citep[][]{ishi08} 
were performed to evaluate their neutrino signal by \citet{sumi09}.  
\citet{self10} studied further the neutrino signal 
utilizing the EOS with quarks and pions \citep[][]{self08a}. Since these exotic constituents soften the EOS and reduce the maximum mass of neutron stars, the time interval between the bounce and black hole formation, which corresponds to the duration of neutrino emission, gets shorter. Therefore the neutrino signal can be used to diagnose the emergence of exotic matter \citep[][]{self09}.

We carry forward here the investigation of the failed supernovae and prompt black hole formations 
with hyperons by paying attention to the ambiguity of hyperon interaction.  
The hyperon interactions with nucleons are relatively well known comparing with other exotic constituents (kaons, pions and quarks). In particular, $\Lambda$-$N$ interaction is well investigated through the single-particle energies of $\Lambda$ hypernuclei. Unfortunately, however, $\Sigma$-$N$ interaction, which affects the components of dense matter and the stiffness of EOS, has a large uncertainity even at present. 
Within this $\Sigma$-$N$ ambiguity, \citet{ishi08} provided several sets of EOS tables of nuclear matter including hyperons. 
In this study, we utilize the EOS tables by \citet{ishi08} to examine both of attractive and repulsive cases for the undetermined $\Sigma$-$N$ interaction supplementing the repulsive case reported 
in the Letter article \citep[][]{sumi09}. 
These Ishizuka EOS tables are based on an $SU_f$(3) extended relativistic mean field (RMF) model and constructed as an extension of the EOS by \citet{shen98a,shen98b}, which is widely utilized in various astrophysical simulations so far. 
Moreover, \citet{ishi08} prepared the hyperon EOS tables including the contributions of thermal pions for the extreme case where an effective mass of pions is equal to their rest mass in vacuum. 
We newly investigate the cases where both of hyperons and pions appear in failed supernovae, 
while the emergence of pions in nucleonic matter was already discussed \citep[][]{self10}.

Therefore, the purpose of this paper is to report the detailed behavior of the black hole formation 
and the neutrino emission with the EOS with hyperons and/or pions.  
We assess especially the dependences on the hyperon-nucleon interaction and pion appearance in hyperonic matter for the black hole formation due to failed supernovae.
We perform numerical simulations of the gravitational collapse, the core bounce, and the following evolution of proto--neutron star up to the black hole formation \citep[][]{sumi07,sumi08}. The general relativistic neutrino radiation hydrodynamics code, which solves the Boltzmann equations for neutrinos together with the Lagrangian hydrodynamics under spherical symmetry, is utilized to compute the dynamics as well as the neutrino signals \citep[][]{yamada97, yamada99, sumi05}. The progenitor model with $40M_{\odot}$ by \citet{woosley95} is adopted as the initial condition for the dynamical simulations. 

This paper is organized as follows. In \S~\ref{numerics} we briefly describe our core collapse model including the numerics and initial setting. In \S~\ref{eos} we explain the hyperonic EOS. The results are discussed in \S~\ref{result}, where we compare the dynamical features and neutrino signals with different hyperonic EOS's. \S~\ref{concl} is devoted to summary and discussion.

\section{Numerical Simulations} \label{numerics}

The numerical simulations are performed with the general relativistic neutrino radiation hydrodynamics code, which solves the Boltzmann equations for neutrinos together with the Lagrangian hydrodynamics under spherical symmetry \citep[][]{yamada97, yamada99, sumi05}. We consider four species of neutrino, $\nu_e$, $\bar \nu_e$, $\nu_\mu$ and $\bar \nu_\mu$, assuming that the distribution function of $\nu_\tau$ ($\bar \nu_\tau$) is equal to that of $\nu_\mu$ ($\bar \nu_\mu$). A detailed description of the numerical simulations such as general relativistic hydrodynamics, transport and reaction rates of neutrinos, and resolutions can be found in \citet{sumi07}. We follow the dynamics from the onset of gravitational collapse of a progenitor through the core bounce and the post-bounce evolution of proto--neutron star by the accretion of outer layer, up to the formation of black hole. Note that we identify the black hole formation by finding the apparent horizon, as explained in \citet{self06}. 
We evaluate the neutrino fluxes and spectra in detail up to the black hole formation owing to the exact treatment of neutrino transfer with a fully implicit method 
together with hydrodynamics. 
Note that, the neutrino-hyperon reactions are omitted for simplicity, which will be revisited later.

As an initial model, we employ the progenitor model of a $40M_{\odot}$ star of \citet{woosley95}. This model contains an iron core of $2.0M_{\odot}$.  We use the profile of its central part up to $3.0M_{\odot}$. While massive stars are suggested to lose their mass during the quasistatic evolutions, mass loss is not taken into account in the progenitor model we adopted. 
The mass loss rate is still uncertain in the theory of stellar evolution.  
The mass loss may affect dynamics and neutrino signals of failed supernovae through the density profile of the outer layer \citep[][]{sumi08,fischer09} as well as convection. 
Recently, \citet{ocon10} performed the core collapse simulations for several sets of the progenitor model with simplified neutrino transfer to find that the outcome depends severely on the mass loss prescription. Since our goal in this study is to investigate the influence of hyperons on the black hole formation, 
we fix the initial condition to a single progenitor model.

\section{Equation of State for Hyperon Matter} \label{eos}

The main input for the numerical simulations of the core collapse of massive stars is 
the EOS of dense matter in addition to the neutrino reaction rates because the dynamical time scale is long enough for particles other than neutrinos to equilibrate. One of the most widely used EOS tables is based on a Skyrme type (non-relativistic) mean field and the liquid-drop model of nuclei \citep[][]{lati91}. Another EOS table is based on an RMF model \citep[][]{suga94}, and nuclear formation is described in the Thomas-Fermi approximation \citep[][]{shen98a,shen98b}. Hyperons are not included in these EOS tables, so that their applicable range may be limited to relatively low temperature and density regions, where hyperons do not emerge abundantly. Recently, several sets of EOS tables of nuclear matter including hyperons using an $SU_f$(3) extended RMF model with a wide range of density, temperature, and charge fraction for numerical simulations of high energy astrophysical processes are presented \citep[][]{ishi08}.

We adopt the 
Ishizuka EOS in this study. This set of EOS is constructed as an extension of Shen EOS and smoothly connected with Shen EOS at low densities. 
In order to include hyperons, they take into account the potential depths of hyperons in symmetric nuclear matter at saturation density $\rho_0$, which are suggested in recent hypernuclear experiments. The potential depth of the $\Lambda$ is well estimated as $U^{(N)}_\Lambda(\rho_0) \sim -30$~MeV from the single-particle energies of many $\Lambda$ hypernuclei. From the recently observed quasi-free $\Sigma$ production spectra, it is considered that $\Sigma$ hyperons would feel a repulsive potential in nuclear matter, $U^{(N)}_\Sigma(\rho_0) \sim +30$~MeV \citep[e.g.,][]{harada05,harada06}. 
Nevertheless, having uncertainties of the potential value, \citet{ishi08} provided several sets of EOS tables using $U^{(N)}_\Sigma(\rho_0) = -30$, 0 $+30$, $+90$~MeV. Also for $\Xi$ hyperons, the analyses of the twin hypernuclear formation \citep[][]{aoki95} and the $\Xi$ production spectra \citep[][]{khaus00} favor a potential depth of around $U^{(N)}_\Xi(\rho_0) \sim -15$~MeV. Note that more accurate measurement for the $\Xi$ potential is expected in the experiment at J-PARC.

The $\Sigma$ and $\Xi$ hyperons are particularly important in neutron stars, since nuclear matter can take a large energy gain from neutron Fermi energy and symmetry energy by replacing, for example, two neutrons with a proton and a negatively charged hyperon ($\Sigma^-$ or $\Xi^-$). If we adopt attractive potential for $\Sigma$ hyperons, $\Sigma^-$ would be the first hyperon to appear in neutron stars. With the recently favored potential around $+30$~MeV, $\Sigma$ hyperons feel a repulsive interaction at high densities, and, hence, cannot appear until the baryon mass density $\rho_B \sim 2\times10^{15}$~g~cm$^{-3}$. Instead of $\Sigma^-$ hyperons, $\Xi^-$ hyperons are found to appear at around $\rho_B \sim 7\times10^{14}$~g~cm$^{-3}$ \citep[see Figure~1 of][]{ishi08}. Therefore, we expect that 
the choice of hyperon interactions have an impact also on the dynamics and neutrino signal of failed supernovae. 

In the sets of Ishizuka EOS, 
the EOS tables including thermal pions are also constructed together with hyperon mixture. The thermal pions are treated in the minimum model where their effective mass is assumed to be equal to their rest mass in vacuum. This approach includes a simple $s$-wave Bose-Einstein condensation of pions, which is different from the pion condensation derived from $p$-wave $\pi N$ interaction \citep[e.g.,][]{knhr93}. Since $s$-wave $\pi N$ interaction is repulsive, their effective mass becomes larger than that in vacuum. In this case, the pion population is suppressed. Thus, this EOS set corresponds to an extreme case where pions are overproduced, provided that $p$-wave $\pi N$ attraction is omitted \citep[][]{onsh09}. 
We adopt this set to assess the maximum effect from thermal pions in hyperon mixture.  

Utilizing the set of Ishizuka EOS described above, we explore the influence of hyperon paying 
attention to the hyperon nucleon interaction and pion mixture.  
We investigate both of attractive and repulsive cases for the $\Sigma$ potential. 
We perform the numerical simulation adopting the EOS sets with 
 the potential depths $(U_\Lambda,U_\Sigma,U_\Xi) = (-30$~MeV, $+30$~MeV, $-15$~MeV) for the repulsive case and ($-30$~MeV, $-30$~MeV, $-15$~MeV) for the attractive case. 
We perform also the corresponding cases adopting the hyperon EOS with pions.  
In the following, we refer to the repulsive EOS with pions, repulsive EOS without pions, attractive EOS with pions and attractive EOS without pions as RP, R, AP and A, respectively. Note that results for the model R were already reported as the Letter article \citep[][]{sumi09}. For comparison, we also show the results for the EOS with pions and without hyperons and Shen EOS (purely nucleonic model), which are referred as NP and N, respectively. The model N was studied by \citet{sumi07} in detail and the model NP was shown also in \citet{self10}, where pions are treated with the same method as in Ishizaka EOS.

The maximum mass of neutron stars gets generally lower due to new hadronic degrees of freedom. Hyperons are no exception. In fact, while the maximum masses of models N and NP are $2.2M_\odot$ and $2.0M_\odot$, respectively, those of every hyperonic models (RP, R, AP and A) are around $1.6M_\odot$. Recently, the mass of the binary millisecond pulsar J1614-2230 was evaluated as $1.97 \pm 0.04M_\odot$ \citep[][]{demo10}. This remarkable precision thanks to a strong Shapiro delay signature excludes the hyperonic models we adopted. 
While many calculations of the zero-temperature EOS for nuclear matter including hyperons were performed in various approaches, almost all of them cannot satisfy such a large value of the maximum mass. At present, Ishizuka EOS is a very limited model including hyperons which is available for astrophysical numerical simulations. 
We utilize Ishizuka EOS in this paper, where our focus is on investigating systematic differences due to the hyperon interaction and pion emergence. 
We believe that the systematics discussed in this study holds for more sophisticated EOS's which will be hopefully free from the maximum mass problem. Note that, very recently, Shen EOS was updated by the authors and $\Lambda$ hyperon is taken into account \citep[][]{shen11}. Their EOS does not include $\Sigma$ and $\Xi$, and its maximum mass is 1.8$M_\odot$.

\section{Numerical Results} \label{result}

\begin{figure}
\plotone{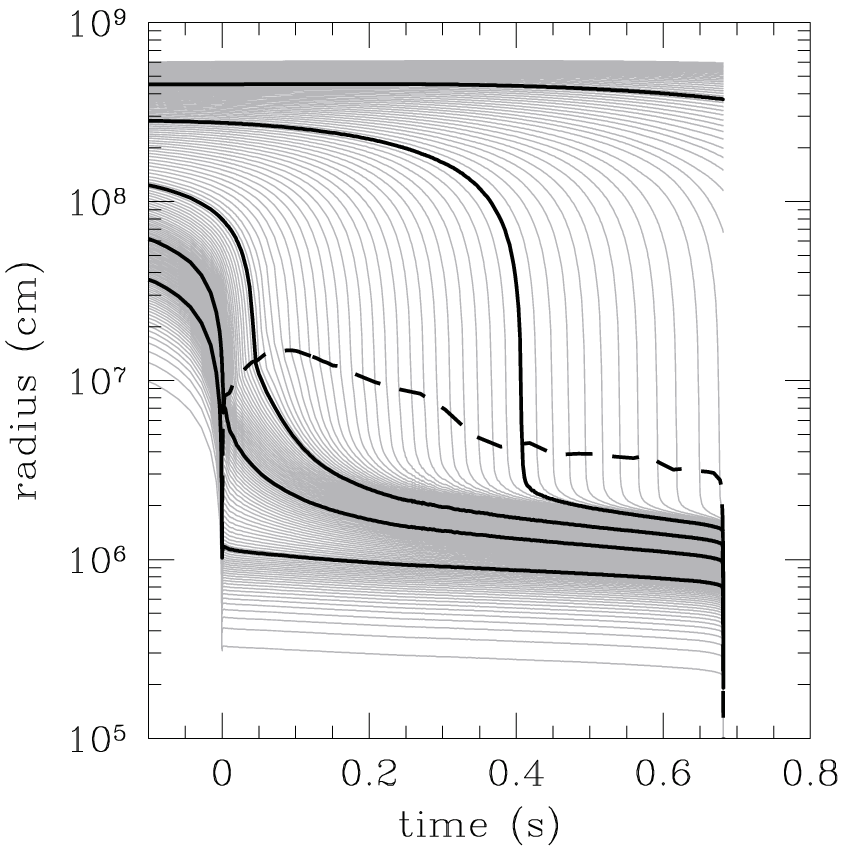}
\caption{Radial trajectories of mass elements as a function of time after bounce in the model with EOS~R. The trajectories are plotted for each $0.02M_\odot$ in baryon mass coordinates. Thick solid lines denote the trajectories for 0.5, 1.0, 1.5, 2.0 and $2.5M_\odot$. The location of the shock wave is shown by a thick dashed line.}
\label{tr}
\end{figure}

To begin with, we describe an outline of the core collapse evolution 
toward the black hole formation using the hyperon EOS. 
The radial trajectories of mass elements in the model with EOS~R are shown in Figure~\ref{tr}. 
The collapse is followed by the core bounce 
due to the nuclear repulsion as in ordinary core collapse supernovae. 
We measure the time from the timing of the core bounce thereafter.  
The central density at the bounce ($3.2 \times 10^{14}$~g~cm$^{-3}$) does not differ among the models very much because the contribution of hyperons and pions is not dominant at this time as revisited later. The shock wave is launched due to the bounce. The propagation of the shock wave is also illustrated in Figure~\ref{tr}. We can see that the shock wave stalls within 100~msec and recedes toward the surface of the central object, proto--neutron star under intense accretion. 
In the meantime, the proto--neutron star contracts gradually 
due to the increase of mass.  
It recollapses to the black hole abruptly at the critical mass. 
These pictures are the same for all the computed models.

\begin{figure}
\plotone{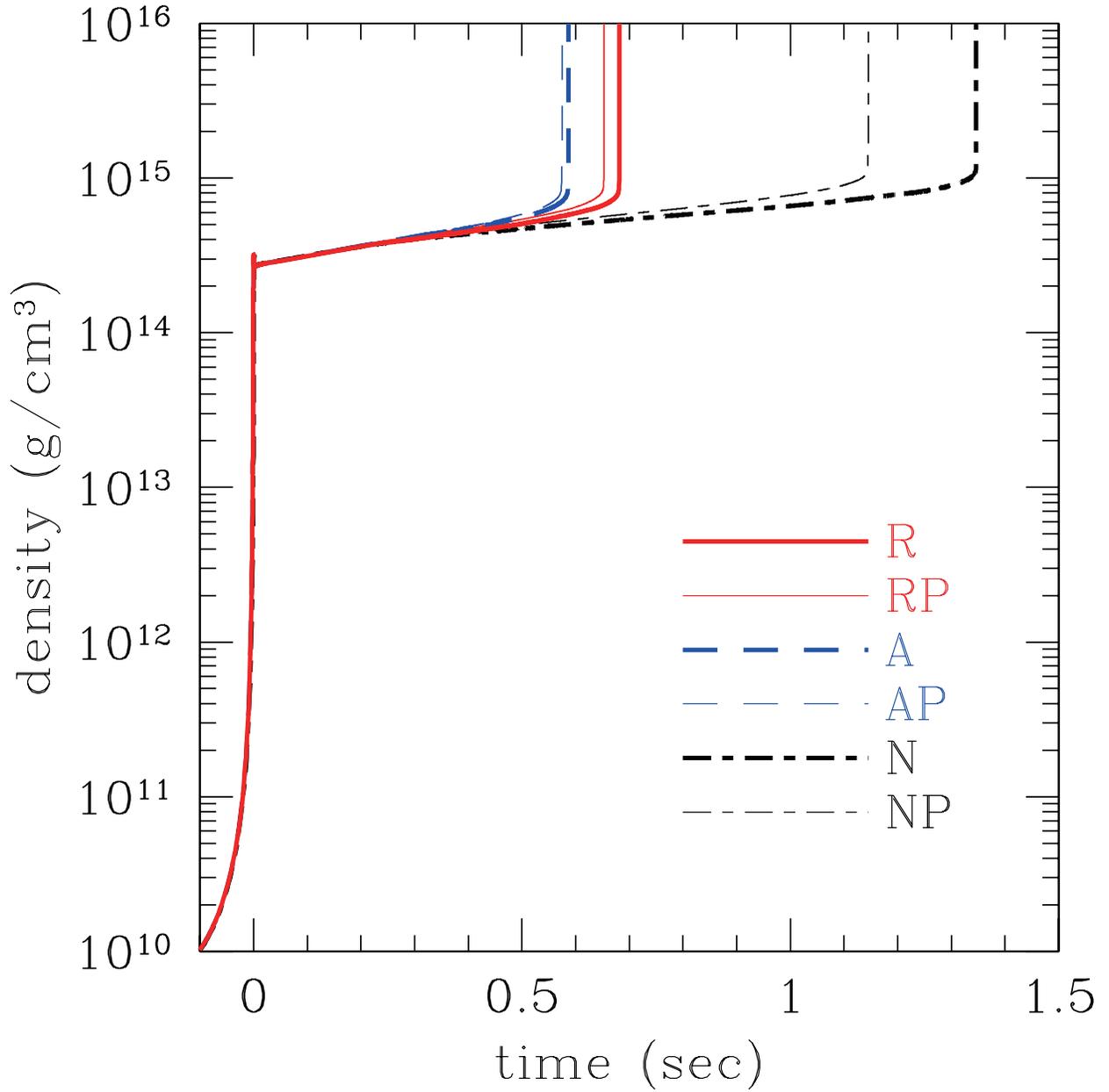}
\caption{Time profiles of the central baryon mass density. Thick solid, thin solid, thick dashed, thin dashed, thick dot-dashed and thin dot-dashed lines correspond to the results for EOS's~R, RP, A, AP, N and NP, respectively.}
\label{cd}
\end{figure}

In Figure~\ref{cd}, we compare the time profiles of the central baryon mass density for all six models. The bounce corresponds to the spikes at $t=0$ and the black hole formation corresponds to the blow-ups. We can recognize that the time interval between the bounce and black hole formation gets shorter 
as we put additional
degrees of freedom, hyperons and pions. 
The time interval 
is 682~msec, 653~msec, 587~msec, 575~msec, 1345~msec and 1145~msec for the models with EOS's~R, RP, A, AP, N and NP, respectively. The black hole formation occurs when the mass of proto--neutron star exceeds its critical mass due to the accretion of outer layer.
Although the proto--neutron star is hot and lepton-rich, 
this critical mass is related with the maximum mass of neutron stars. As already mentioned, new hadronic degrees of freedom decrease the maximum mass of neutron stars. On the other hand, the mass accretion rate does not differ among the EOS models because hyperons and pions do not appear in the outer layer with low density and temperature. 
Hence the inclusion of hyperons and pions simply hasten the black hole formation. In contrast, the initial location of the apparent horizon, which is typically 1.1-1.2$M_\odot$ in the baryon mass coordinate, does not depend on the EOS.

\begin{figure}
\plotone{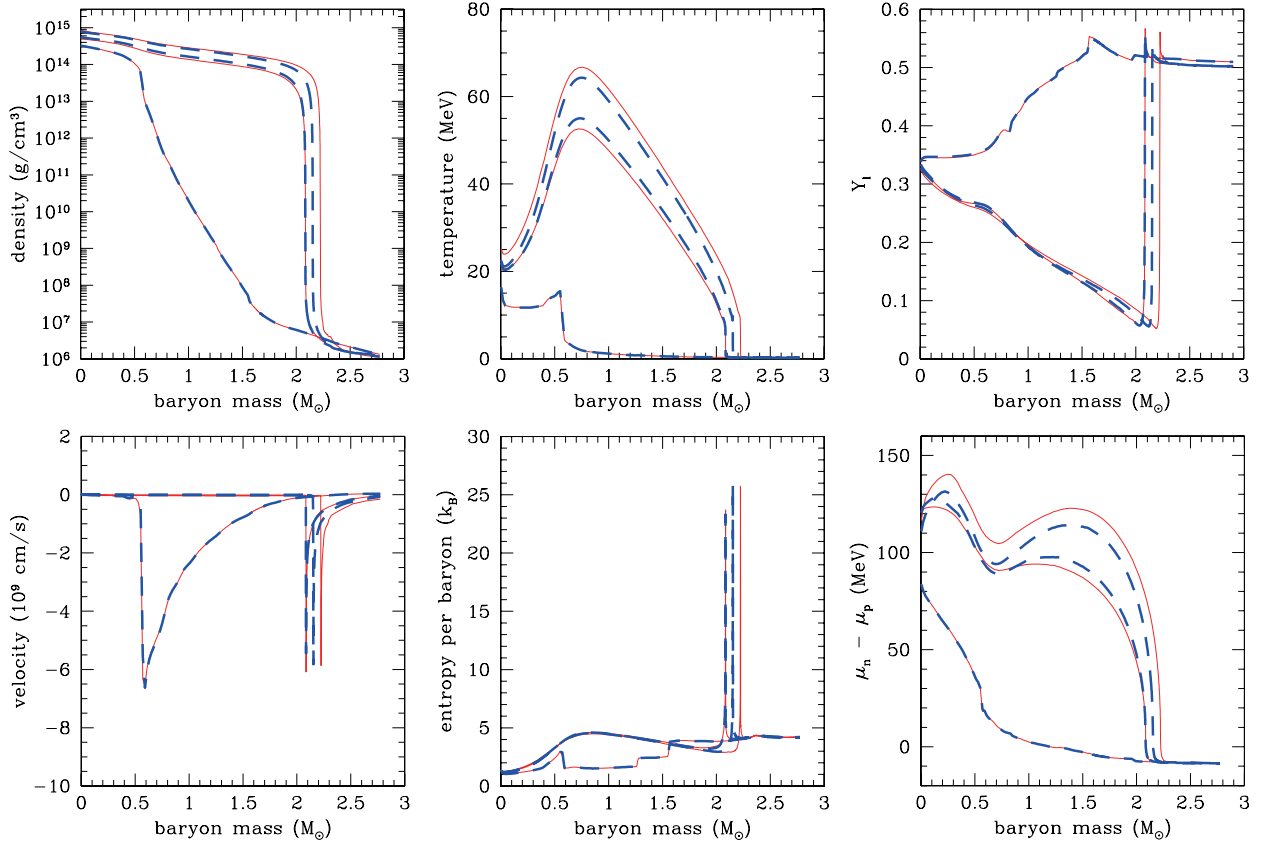}
\caption{Profiles of the density (upper left), temperature (upper center), electron-type lepton fraction (upper right), velocity (lower left), entropy per baryon (lower center) and charge chemical potential $\mu_n - \mu_p$ (lower right) at bounce, 500~msec after bounce and 2~msec before black hole formation. The results for EOS's~R and A are shown by thin solid and thick dashed lines, respectively. Note that the 2~msec before black hole formation corresponds to 680~msec after bounce for EOS~R and 585~msec after bounce for EOS~A.}
\label{tp}
\end{figure}

\begin{figure}
\plotone{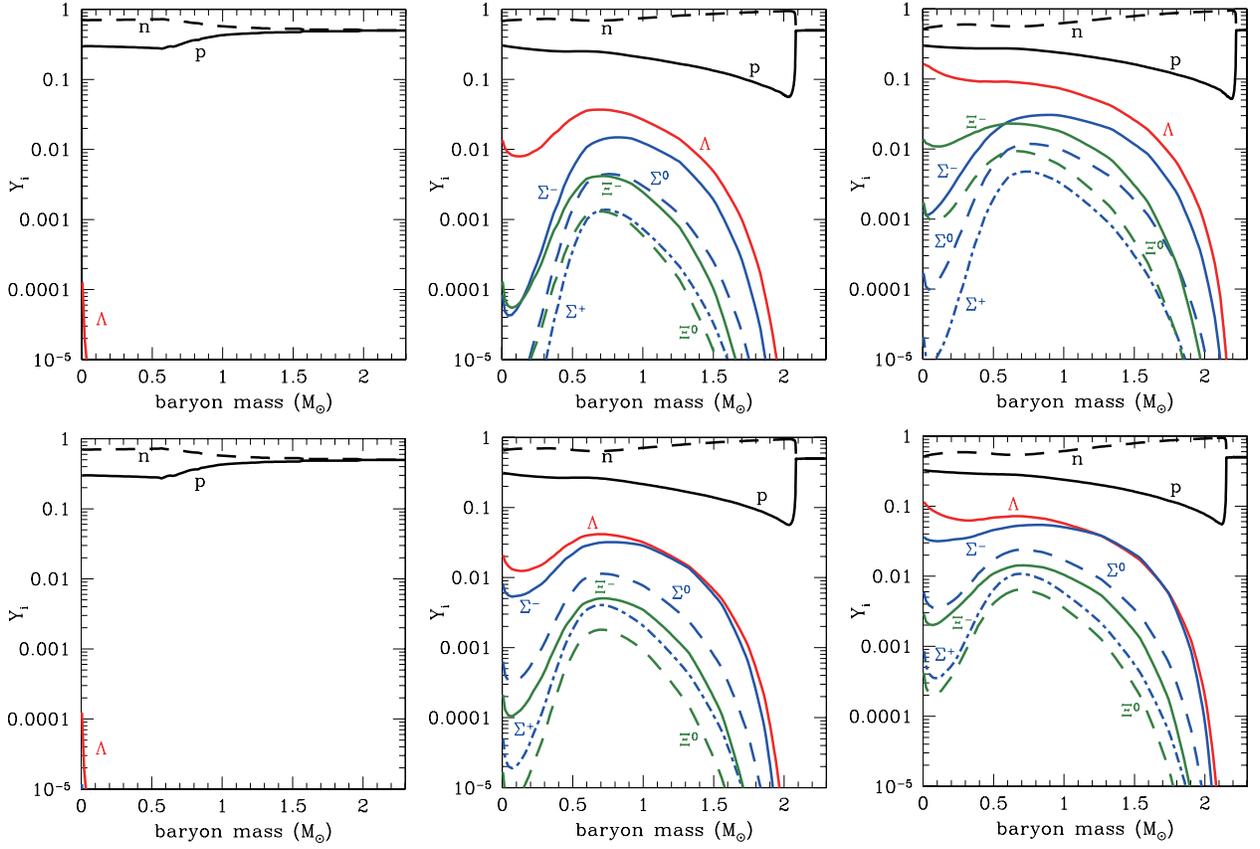}
\caption{Profiles of the particle fractions at bounce (left), 500~msec after bounce (center) and 2~msec before black hole formation (right). The upper and lower panels show results for EOS's~R and A, respectively.}
\label{yi}
\end{figure}

The neutrino duration time during black hole formation of the model with EOS~R is $\sim$15\% longer than that of the model with EOS~A. Since the maximum masses of EOS's~R and A do not differ very much, the duration time is more sensitive to the hyperonic matter EOS than the neutron star maximum mass is. Here
we examine the profiles of the models with EOS's~R and A to discuss 
the differences due to the hyperon interaction in the black hole formation. The profiles of key quantities at the selected times are shown in Figure~\ref{tp}. 
At the core bounce, 
we can see that the hyperon potential does not make any difference because hyperons do not emerge yet. This is also evident from Figure~\ref{yi}, where the profiles of particle fractions are shown. After the bounce, the density and temperature rise due to the contraction. This process is adiabatic as can be recognized from the entropy profiles. The electron-type lepton fraction does not also vary 
very much during this stage.

Comparing the profiles at 500~msec after bounce, we can see that density and temperature of the model with EOS~A are somewhat higher than that of EOS~R while locations of the shock are the same for both models (see the velocity profile). As seen in Figure~\ref{yi}, $\Lambda$ and $\Sigma^-$ hyperons are populated near the center for the model with EOS~A. By contrast, for the model with EOS~R, the appearance of $\Sigma^-$ hyperons is suppressed due to the repulsive potential. Therefore the density and temperature get higher for the attractive case because the EOS is more softened by the effect of hyperons. This is also the reason why the time interval between the bounce and black hole formation is shorter for the model with EOS~A comparing with the model with EOS~R.

\begin{figure}
\plotone{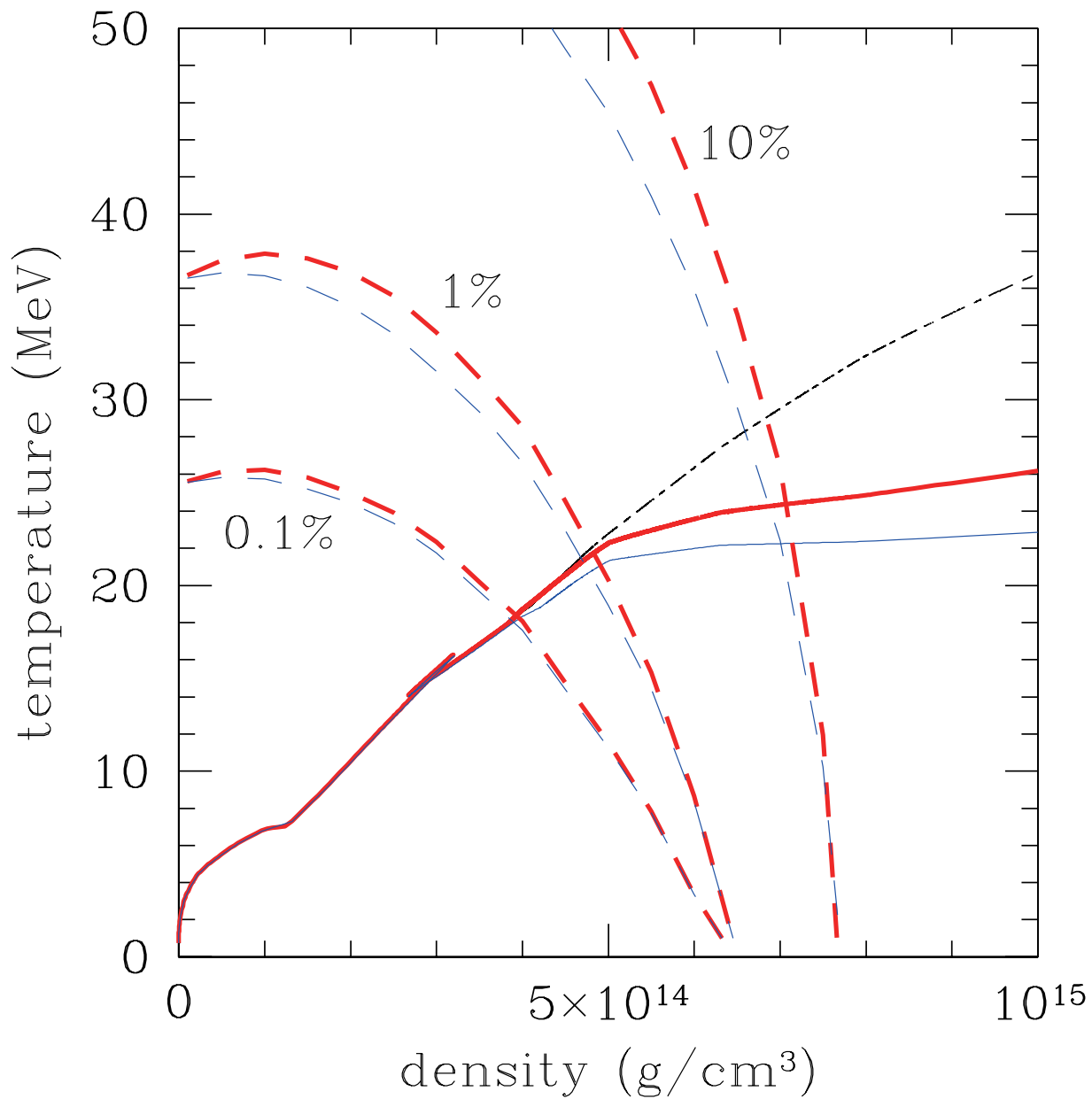}
\caption{Evolutions of the central density and temperature (solid lines) with hyperon fraction contours for the electron fraction $Y_e=0.3$ (dashed lines). Thick and thin lines correspond to EOS's~R and A, respectively. The dot-dashed line shows the trajectory of the model with EOS~N.}
\label{dt}
\end{figure}

We discuss here the profiles at 2~msec before the black hole formation. Note that this moment corresponds to 680~msec after bounce for EOS~R and 585~msec after bounce for EOS~A. The central temperature of EOS~A is lower than that of EOS~R whereas their central densities are almost the same. This feature can be also recognized from Figure~\ref{dt}, where the evolutions of the central densities and temperatures are shown. As already mentioned, evolutionary tracks are roughly isentropic lines. Therefore EOS~R has higher temperature than EOS~A for the same value of the entropy. This is equivalent to the fact 
that EOS~A has higher entropy than EOS~R for the same value of the temperature. In the attractive case, since $\Sigma$ hyperons appear more easily and the number of particle species increases, the entropy gets higher for the fixed temperature.

The peak of temperature profile resides not at the center but at the medium region ($\sim$0.7$M_\odot$ of the baryon mass coordinate) for both models. This is because the entropy becomes high 
due to the shock heating at the medium region. The hyperon fractions are also different between the central and medium regions. For the repulsive case, $\Xi^-$ is the dominant negatively charged hyperon at the center, as is the case for cold neutron stars. In the medium region, the fraction of $\Sigma^-$ is larger than that of $\Xi^-$ because of the thermal population. On the other hand, for the attractive case, $\Sigma^-$ appears more than $\Xi^-$ in the entire region. The abundance of negatively charged hyperons is also reflected in the charge chemical potential, which is the difference between the chemical potential of neutrons and that of protons, $\mu_n - \mu_p$. For the attractive case, the charge chemical potential is reduced owing to the emergence of $\Sigma^-$.

\begin{figure}
\plotone{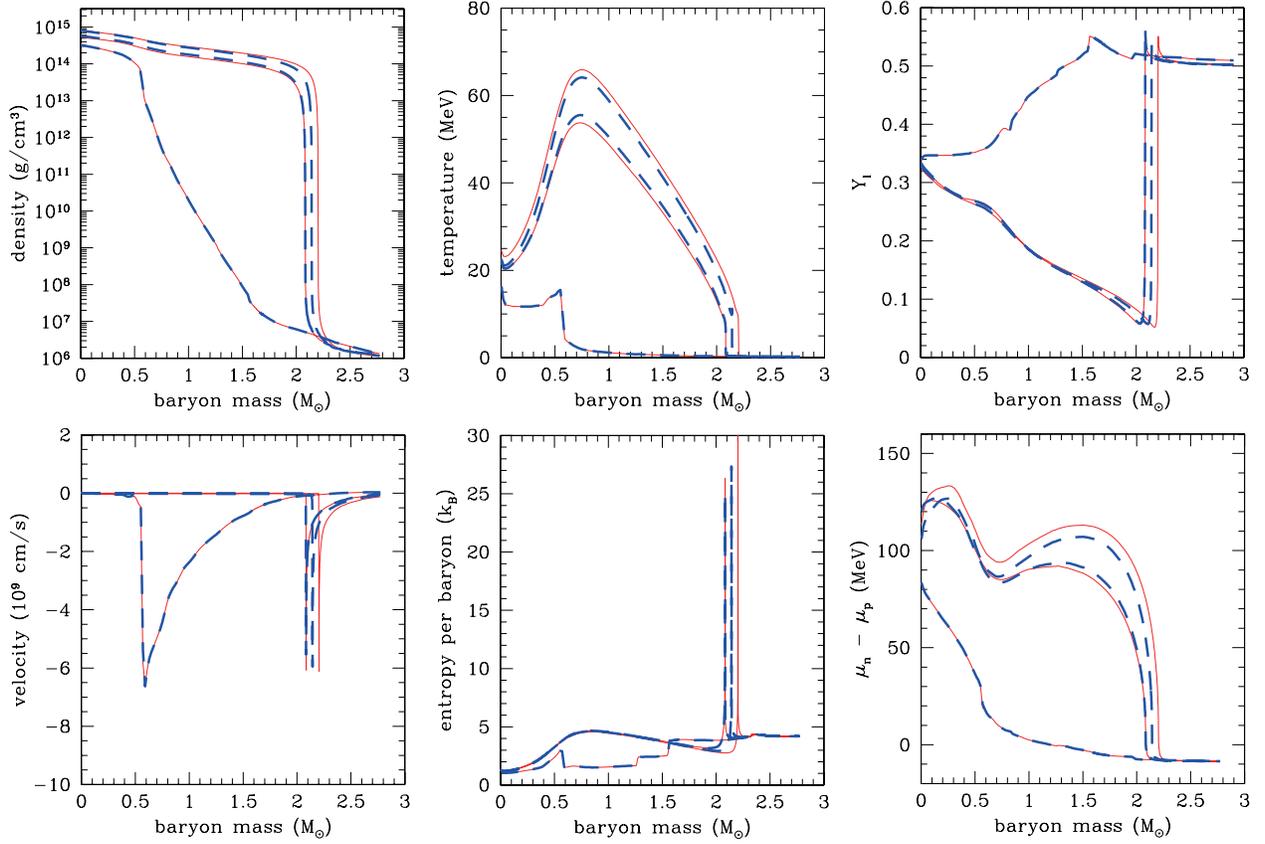}
\caption{Same as Figure~\ref{tp} but for EOS's~RP (thin solid lines) and AP (thick dashed lines). Note that the 2~msec before black hole formation corresponds to 651~msec after bounce for EOS~RP and 573~msec after bounce for EOS~AP.}
\label{tpp}
\end{figure}

We now turn to the effect of pions. As shown in Figure~\ref{tpp}, qualitative features of the key quantities for EOS's~RP and AP are similar to their counterparts without pions. Since the EOS is softened by the contribution of pions, the compression is accelerated both of the repulsive and attractive cases. As a consequence, the difference between the models with EOS's~RP and AP is more minor than that of EOS's~R and A. The appearance of pions results in the decrease of the electron fraction because the negative charge is shared by electrons and $\pi^-$. An influence of pions is more clearly seen in the profiles of the charge chemical potential. The chemical potential of charged pions, $\mu_{\pi^-}$, is equivalent to the charge chemical potential due to the balance, $n \leftrightarrow p + \pi^-$. On the other hand, since the pions are thermally populated in our model, $\mu_{\pi^-}$ is limited by their rest mass in vacuum, $m_{\pi^-}=135$~MeV. 
We can see $\mu_{\pi^-}$ saturates near $0.2M_\odot$ 
from the profile of the charge chemical potential of the model with EOS~RP at 2~msec before black hole formation.

\begin{figure}
\plotone{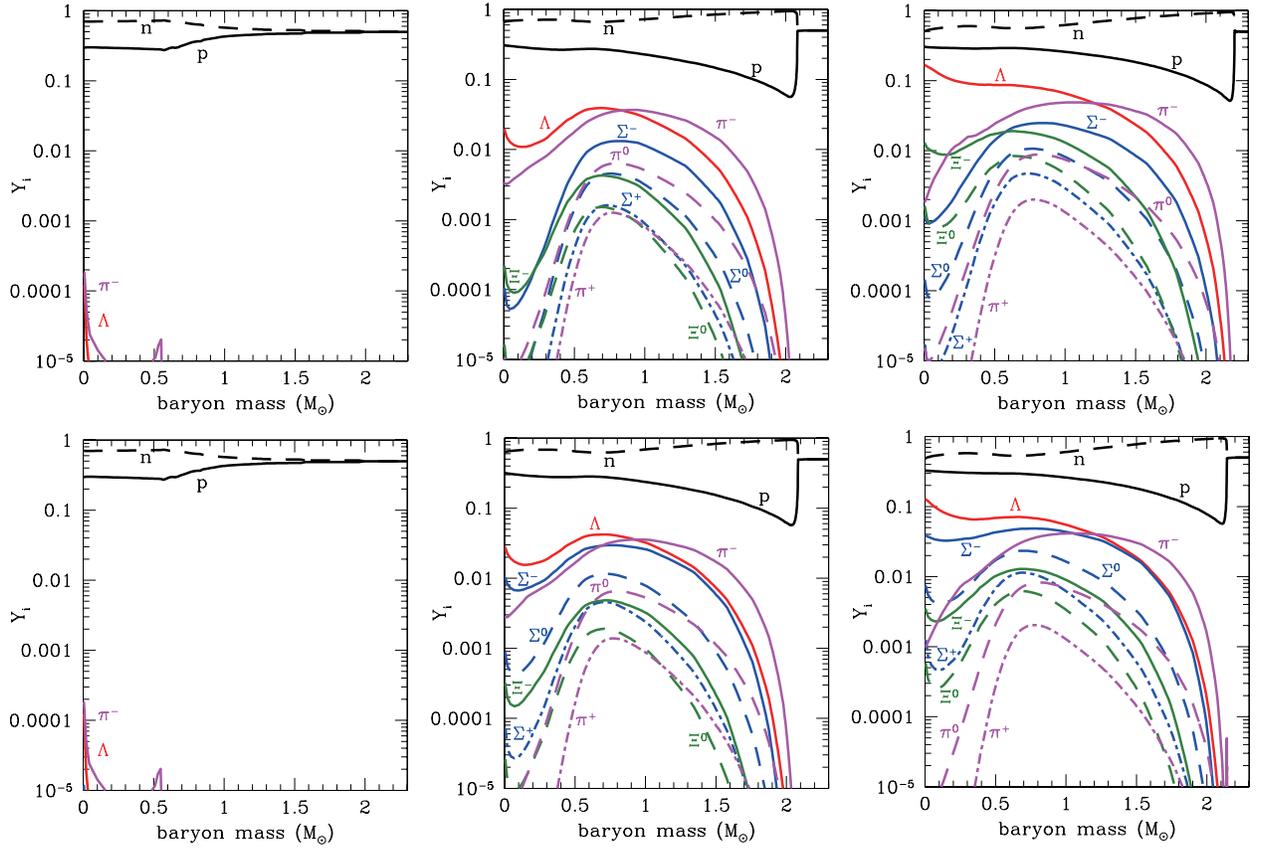}
\caption{Same as Figure~\ref{yi} but for EOS's~RP (upper panels) and AP (lower panels).}
\label{yip}
\end{figure}

The profiles of particle fractions with pions are shown in Figure~\ref{yip}. While there are quantitative differences between the models with and without pions, the pion mixing does not change the order of fractions for each hyperon. The fraction of $\pi^-$ at the center is larger for 500~ms after bounce than that for 2~msec before black hole formation because pions are replaced by the negatively charged hyperons for high density at 2~msec. During the contraction, pions appear so as to decrease the charge chemical potential. On the other hand, the octet baryons should be populated equally in high density or temperature limit, where the charge neutrality is already satisfied and pions are not necessary. Therefore, the existence of pions is limited within the intermediate density regime.

\begin{figure}
\plotone{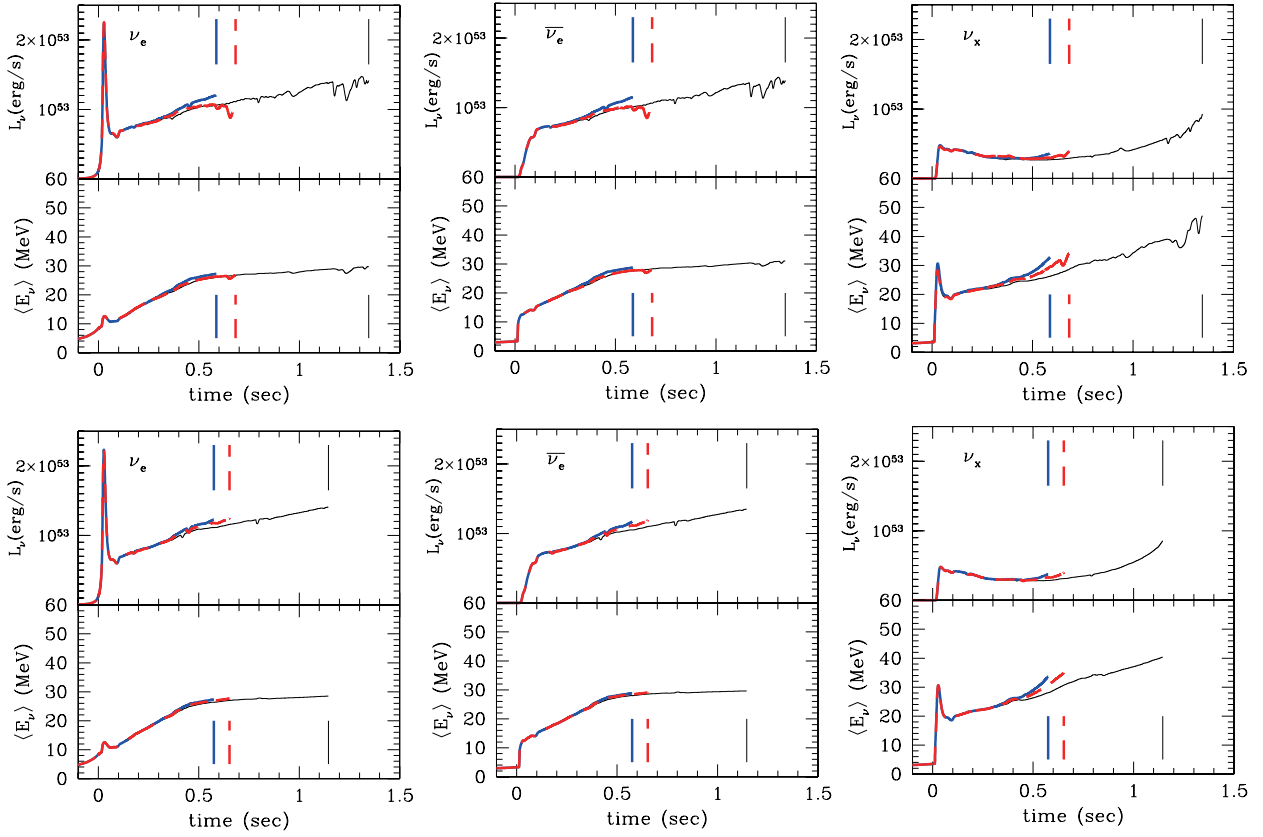}
\caption{Luminosities (upper plots) and average energies (lower plots) of the emitted neutrinos as a function of time after bounce. The panels correspond, from left to right, to $\nu_e$, $\bar\nu_e$ and $\nu_x$ ($=\nu_\mu$, $\nu_\tau$, $\bar\nu_\mu$, $\bar\nu_\tau$). The results for the models with EOS's~R (thick dashed lines), A (thick solid lines) and N (thin solid lines) are shown in the top panels, and the results for the models with EOS's~RP (thick dashed lines), AP (thick solid lines) and NP (thin solid lines) are shown in the bottom panels. Vertical lines represent the end point of the neutrino emission.}
\label{nus}
\end{figure}

The core collapse of massive stars is accompanied by the emission of a large amount of neutrinos. The time profiles of luminosities and average energies of emitted neutrinos are shown in Figure~\ref{nus} for all models. 
Note that $\nu_\mu$ and $\bar\nu_\mu$ have the same type of reactions, the difference in coupling constants is minor and $\nu_\tau$ ($\bar\nu_\tau$) is assumed to be the same as $\nu_\mu$ ($\bar\nu_\mu$). 
Therefore, we collectively denote these four species as $\nu_x$. The average energy presented here is defined by the root mean square value. The luminosity is corrected by the gravitational red-shift, though it is minor up to the end of our simulation at the apparent horizon formation. At the moment of black hole formation, the neutrino luminosity decreases steeply due to the red-shift within a short period and the emission stops suddenly \citep[e.g.,][]{baum96,beacom01}\footnote{However see also \citet{ss11} for moderately rotating collapse.}. Therefore the time interval between the bounce and black hole formation almost corresponds to the duration of neutrino emission. As seen in the figure, the duration of neutrino emission is different among the models. This fact implies that the neutrino signal as an observable reflects the difference in the hyperon-nucleon interaction through the black hole formation.

The neutrino signal of the stellar collapse can be seen as follows. The neutrino optical depth $d(r)$ at the radius $r$ is defined as
\begin{equation}
d(r) = \int^{R_s}_{r} \frac{dr^{\prime}}{l_\mathrm{mfp}(r^{\prime})},
\label{depth}
\end{equation}
where $R_s$ is the stellar radius and $l_\mathrm{mfp}$ is the mean free path of neutrino, and the radius of the neutrino sphere $R_\nu$ is defined as
\begin{equation}
d(R_\nu) = \frac{2}{3},
\label{sphere}
\end{equation}
for neutrinos with a typical energy. Roughly speaking, neutrinos are emitted from the neutrino sphere. In our cases, hyperons and pions appear only for the central region where the density and temperature are high enough for neutrinos to be trapped, i.e., $d(r) \gg 1$. In fact, 
the time evolutions of the neutrino signals are quite similar among the models 
except for the end points.  
We can recognize that the neutrino luminosity and average energy are slightly higher for the models with hyperons and/or pions. This is because they affect the location and temperature of the neutrino sphere through the softening of EOS. 

The neutrino luminosity summed over all species is given approximately by the accretion luminosity \citep[][]{thomp03}
\begin{equation}
L^\mathrm{acc}_\nu \sim \frac{GM_\nu \dot{M}}{R_\nu},
\label{acclumi}
\end{equation}
 where $G$, $\dot{M}$ and $M_\nu$ are the gravitational constant, the mass accretion rate and the mass enclosed by $R_\nu$, respectively. In our case, as is the case for ordinary supernovae, the proto--neutron star is formed before the black hole formation, and neutrinos are emitted mainly on the surface of the proto--neutron star. Thus we can regard $R_\nu$ and $M_\nu$ as the radius and mass of the proto--neutron star, respectively. The radius of the proto--neutron star becomes smaller for the models with hyperons and/or pions because of the softening of EOS. Thus the neutrino luminosity gets higher. On the other hand, the average energy of neutrinos is approximately proportional to the temperature of the neutrino sphere, $T_\nu$. When $R_\nu$ decreases because of the softening of EOS, $T_\nu$ increases. This is the reason why the average energy of neutrinos gets higher for the models with  hyperons and/or pions. These features are remarkable for the case with an attractive $\Sigma$-$N$ interaction because $\Sigma^-$ hyperons are easy to appear and help the softening of EOS. Incidentally, in order to examine these differences in the neutrino signals using actual events of failed supernovae, more detailed statistical analyses is necessary based on the method in \citet{self09}. Progenitors in our Galaxy will be favorable for the statistical studies.

\begin{figure}
\plotone{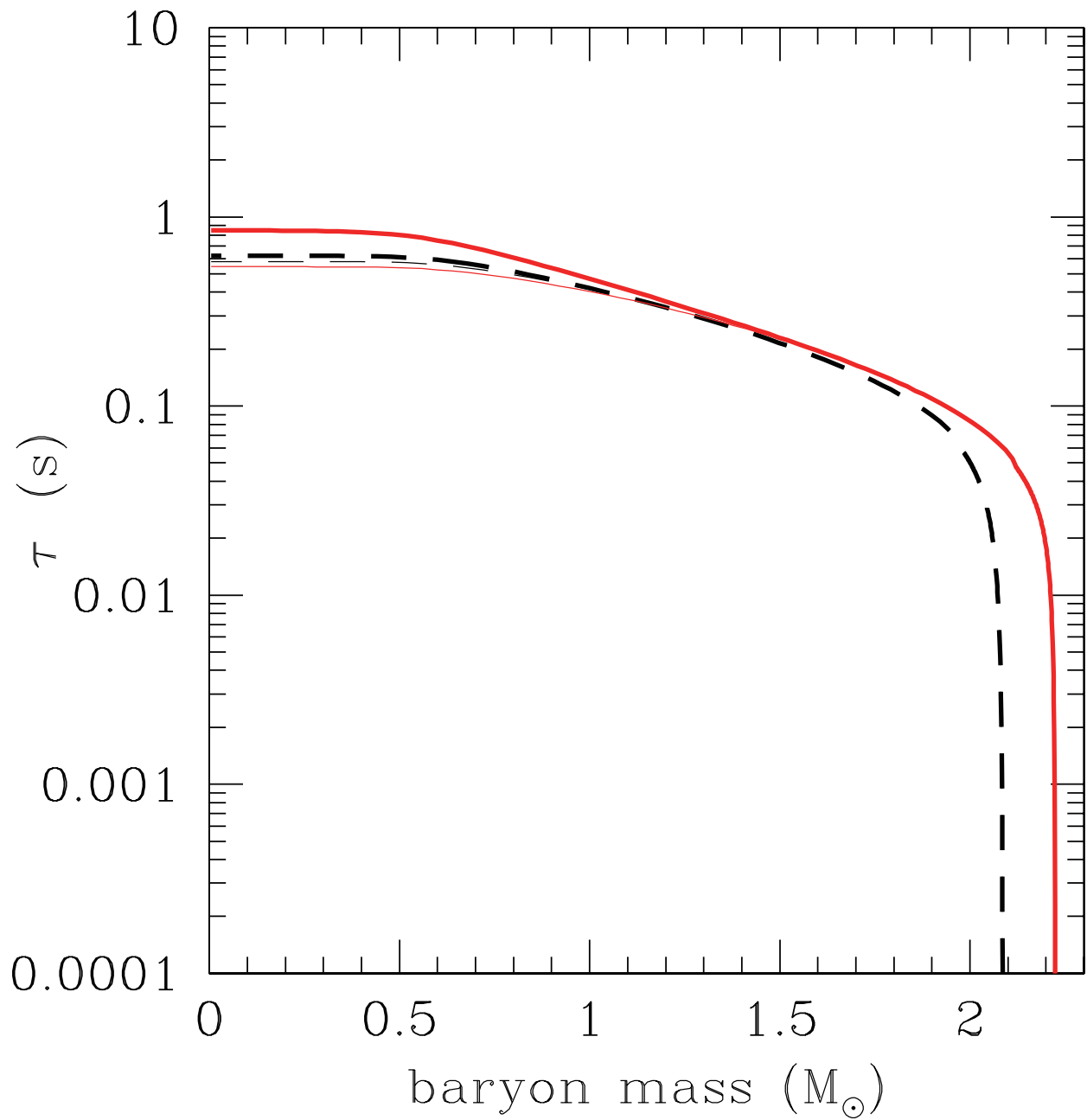}
\caption{Profiles of the diffusion time of $\nu_e$ with 30~MeV for the model~R at 500~msec after bounce (dashed lines) and 680~msec after bounce (solid lines). Thick lines represent cases with the neutrino-hyperon reactions while thin lines represent cases without the neutrino-hyperon reactions.}
\label{furu}
\end{figure}

We have checked that the omission of neutrino-hyperon reactions in our simulation 
is a reasonable approximation through the evaluation of diffusion time scale.  
In fact, hyperons appear only in the central region and neutrinos in the central region take time to diffuse out. Thus the absence of the neutrino-hyperon reactions will not make a difference for the early phase. The diffusion time from the radius $r$ can be evaluated as
\begin{equation}
\tau(r) = \int^{R_s}_{r} \frac{6r^{\prime} dr^{\prime}}{c \, l_\mathrm{mfp}(r^{\prime})},
\label{diffuse}
\end{equation}
where $c$ is the velocity of light. In Figure~\ref{furu}, we compare the diffusion time of $\nu_e$ with 30~MeV, which is typical energy of emitted neutrinos, of the model~R for the cases with and without the neutrino-hyperon reactions based on the reaction rates shown in \citet{reddy98}. We can recognize that the difference is subtle for 500~msec after bounce because the hyperon fraction is tiny. Since the difference resides inside $1M_\odot$, where the diffusion time is $\gtrsim$500~ms, the influence of hyperons will be seen in the neutrino signals after $\gtrsim$500~ms disregarding the evolution of the mean free path. At 2~msec before black hole formation, neutrino-hyperon reactions makes a noticeable difference inside $1.3M_\odot$, but the diffusion time is still long with $\tau \gtrsim 250$~ms. Therefore we can conclude that the omission of the neutrino-hyperon reactions does not affect the results on the emitted neutrino signals.

\section{Summary and Discussion} \label{concl}
In this paper, we have presented the emergence of hyperons in black-hole-forming failed supernovae
with the focus on the hyperon-nucleon interaction and the pion mixture. 
A series of core-collapse simulation for a progenitor model with $40M_\odot$ have been performed solving general relativistic neutrino radiation hydrodynamics under spherical symmetry. We have adopted a few sets of EOS tables of nuclear matter including hyperons using an $SU_f$(3) extended RMF model. In particular, we have paid attention to the difference in the $\Sigma$ potential, whether attractive or repulsive. The appearance of thermal pions have been also taken into account including a simple $s$-wave Bose-Einstein condensation. The collapse leads to the core bounce and the birth of proto--neutron stars with increasing mass, while the shock wave stalls and the explosion fails being different from ordinary core-collapse supernovae. We have found that the inclusion of hyperons and pions promotes the recollapse towards the black hole formation by terminating the proto--neutron star epoch because the EOS is softened due to new hadronic degrees of freedom. Since $\Sigma$ hyperons appear more easily, the softening is more efficient for the attractive $\Sigma$ potential cases comparing with the repulsive counterparts.  
We have found also that thermal pions act as an additional source of softening together with hyperons.  

We have also evaluated the amount of neutrinos emitted from the failed supernovae. For the spherically symmetric nonrotating collapse, the time interval from the bounce to the black hole formation corresponds to the duration of neutrino emission. 
The softer EOS's with new hadronic degrees of freedom lead to the smaller critical masses and 
the earlier timing of black hole formation.  
Therefore the duration of neutrino signal is shorter for the models with hyperons and/or pions.
The time evolutions of the neutrino signals before the black hole formation are found rather similar 
among the models other than the difference in duration. 
This is because they appear only inside the proto--neutron star, which is opaque to neutrinos. 
However, there are slight differences in the rise of energies and luminosities 
around the end points in the duration of neutrino signal.  
They would provide a clue to assess the physics of strangeness 
in addition to the information from the time duration.  
In fact, the event number of neutrinos at SuperKamiokande detector is evaluated for the Galactic events as $\sim$10$^4$, which could be a target of statistical studies \citep[][]{self09}.

As already mentioned, there are large uncertainties in the mass loss prescription of the progenitor model. It affects the stellar structure and, therefore, the time interval between the bounce and black hole formation \citep[][]{sumi08,fischer09,ocon10}. Thus it is our concern that the difference in the progenitor mass produces changes in the duration of the neutrino signal as the difference in the EOS does. However, there may be the observational chances for probing the progenitor apart from the neutrino. For instance, since we can determine the direction of the progenitor to some extent by the neutrino detection \citep[][]{ando02}, optical follow-up observations may be possible. While the neutrino sphere is swallowed by the black hole for in $\sim$1 sec, the free fall time at the stellar surface (photosphere) is $1/\sqrt{G \rho} \sim 10^7$ sec. By observing the stellar disappearance, we may be able to get more information on the stellar structure. Combining such additional data, we believe that the assessment of the EOS from the neutrino signal is promising.

As a final note, we would like to emphasize that the knowledge about the properties of dense matter at extreme conditions is essential for the studies on the astrophysical black hole formation such as not only the collapse of massive stars but also the merger of binary neutron stars \citep[][]{seki11} and so on.

\acknowledgments

We are grateful to Chikako Ishizuka for fruitful discussions. In this work, numerical computations were partially performed on the supercomputers at Research Center for Nuclear Physics (RCNP) in Osaka University, Center for Computational Astrophysics (CfCA) in the National Astronomical Observatory of Japan (NAOJ), Yukawa Institute for Theoretical Physics (YITP) in Kyoto University, Japan Atomic Energy Agency (JAEA), High Energy Accelerator Research Organization (KEK) and The University of Tokyo. This work was partially supported by Research Fellowship for Young Scientists from the Japan Society for Promotion of Science (JSPS) through 21-1189, and Grants-in-Aids for the Scientific Research (Nos.~19104006, 19540252, 21540281, 22540296, 23840038) and Scientific Research on Innovative Areas (Nos.~20105004, 20105005) from the Ministry of Education, Culture, Sports, Science and Technology (MEXT) in Japan.

%\appendix

\clearpage

\end{document}